\documentclass{phostproc}

\usepackage{lipsum} %%% REMOVE THIS 

\title{PB 8783: the first sdO star suitable for asteroseismic modeling?}
\author{Valerie Van Grootel,$^{1}$ 
        Suzanna K. Randall,$^{2}$ 
        Marilyn Latour,$^{3}$ 
        Peter N\'emeth,$^{4}$ 
        Gilles Fontaine,$^{5}$ 
        Pierre Brassard,$^{5}$ 
        Stephane Charpinet,$^{6}$ 
        Elizabeth M. Green$^{7}$}

\affiliation{$^{1}$ Space sciences, Technologies and Astrophysics Research (STAR) Institute, Universit\' e de Li\` ege, 19C All\'ee du 6 Ao\^ ut, B-4000 Li\` ege, Belgium  \\
			 $^{2}$ ESO, Karl-Schwarzschild-Str. 2, 85748, Garching bei Munchen, Germany \\
			 $^{3}$ Institute for Astrophysics, Georg-August-University, Friedrich-Hund-Platz 1, 37077, Gottingen, Germany \\
			 $^{4}$ Astronomical Institute of the Czech Academy of Sciences, CZ-251 65, Ondrejov, Czech Republic ; Astroserver.org, 8533 Malomsok, Hungary \\
			 $^{5}$ Universit\'e de Montr\'eal, D\'epartement de Physique, C.P. 6128 Succ. Centre-ville, Montr\'eal, QC H3C 3J7, Canada \\
			 $^{6}$ IRAP, Universit\'e de Toulouse, CNRS, UPS, CNES, 14 avenue Edouard Belin, F-31400, Toulouse, France \\
			 $^{7}$ Steward Observatory, University of Arizona, 933 North Cherry Avenue, Tucson, AZ, 85721, USA}

\shorttitle{PB 8783}
\shortauthors{V. Van Grootel et al.}

\abs{Pulsating hot B subdwarfs (sdB) have become one of the showcases of asteroseismology. Thanks to the combination of rich observed pulsation spectra and state-of-the-art modeling tools it is possible to tightly constrain fundamental parameters such as the stellar mass for these compact, evolved objects in the core He-burning phase of evolution. In comparison, the exploitation of sdO pulsators - hotter stars thought to be in a more advanced evolutionary stage - is still in its infancy. While a small number have been identified in Globular Clusters (GCs), they are extremely rare among the field population. It was recently suggested that PB8783, one of the very first sdB pulsators discovered in 1997, may in fact be an unrecognized hot sdO star with very similar properties to the GC sdO pulsators. Here, we present new very high-quality spectroscopy as well as an asteroseismic analysis of this star with the aim of solving the mystery of its nature.}

\begin{document}

\maketitle

\section{Introduction to sdB and sdO stars}
Hot subdwarfs of the B (sdB) and O (sdO) types form the bluer extension of the horizontal branch, also referred to as the extreme horizontal branch (EHB) in the Hertzsprung-Russell diagram. While sdB stars (22 000 K $\lesssim T_{\rm eff} \lesssim$ 38 000 K) correspond to core He-burning objects, their hotter sdO counterparts  (38 000 K $\lesssim T_{\rm eff} \lesssim$ 100 000 K) correspond to various evolutionary channels. Some sdOs are believed to be the direct progeny of sdB stars evolving through a short-lived phase of He-shell burning on their way towards the white dwarf cooling sequence. The H-rich envelope of sdB and sdO stars is indeed too thin to sustain significant H-shell burning, thus these stars do not reach the Asymptotic Giant Branch (AGB). For an extensive review on sdB and sdO stars, see \citet{2016PASP..128h2001H}.

The EHB hosts various classes of pulsating stars. The sdBV class is the oldest, largest (about 100 sdBVs are known at the time of writing, just before the TESS era) and best studied class, and can be subdivided into the short-period pulsators (sdBV$_r$, discovered in 1997) and the long-period pulsators (the sdBV$_s$, discovered in 2003). The short-period pulsations (80-600 s) usually correspond to low- order p-modes, while long-period pulsations (30 min - 3h) are generally mid- to high-order g-modes. Many sdB pulsators are actually hybrid pulsators (as revealed by \textit{Kepler}), with predominantly short-period oscillations, or predominantly long-period oscillations. Their pulsating sdO (sdOV) field counterparts are rare, and have been somewhat desperately searched for over the years (see, e.g., \citealt{2007MNRAS.379.1123R,2014ASPC..481..153J}), following the discovery of the first sdOV by \citet{2006MNRAS.371.1497W}. It is only recently, twenty years after his discovery of the first sdB pulsator in 1997, that \citet{2017MNRAS.467.3963K} announced the second bona-fide field sdO pulsator. In the meantime, a survey dedicated to the search of pulsating EHB stars in the Globular Cluster (GC) $\omega$\,Centauri uncovered five pulsating objects, which spectroscopic follow-up observations revealed to be all hot sdO-type subdwarfs \citep{2016A&A...589A...1R}. Interestingly, no sdB stars have yet been confirmed to pulsate in GCs\footnote{Six pulsating EHB stars have been discovered in NGC\,2808 by \citet{2013ApJ...777L..22B} but unfortunately their spectral type remain uncertain due to a lack of spectroscopic information.}.

To date, 15 sdB pulsators have been modeled by asteroseismology \citep{2012A&A...539A..12F}. Thanks to a forward modeling approach (see Sect. \ref{method}), asteroseismology allows us to access the global ($M_*$, $\log g$, $R_*$, etc.) and structural parameters ($M_{\rm env}$, $M_{\rm core}$, core composition, etc.) of an individual star. It notably helps to clarify the question of the origin of sdB stars, e.g. through their mass distribution that is strongly peaked at $\sim$ 0.47 $M_{\odot}$.  This compares qualitatively very well with the expectation that the majority of sdB stars are post-RBG objects having lost almost all of their H-envelope through binary interaction with stellar, substellar, or maybe planetary companions \citep{2014ASPC..481..229V}. In contrast, none of the known sdO pulsators have been modeled by asteroseismology so far. This is partly due to the fact that it is extremely challenging to obtain the high-quality photometry necessary for seismic modeling for the sdO pulsators known in GCs. In addition, the accurate spectroscopic parameters (such as $T_{\rm eff}$) needed to guide any asteroseismic modeling are difficult to obtain for sdO stars, especially if they are faint (like the $\omega$\,Cen pulsators). Space-based UV spectroscopy can be helpful for constraining the effective temperature of hot sdO stars by using the ionization equilibrium of metal lines (see, e.g., \citealt{2017AJ....154..126D}), but such observations are notoriously difficult to obtain and their analyses can be hampered by severe crowding of the spectral lines \citep{2017A&A...600A.130L}.  
 
PB 8783 is a binary star consisting of a pulsating subdwarf and a main sequence F-type companion. It was actually one of the very first pulsating subdwarfs discovered \citep{1997MNRAS.285..645K} and was thought to be a bonafide sdBV$_r$ pulsator for years. However, since the F-companion dominates the flux in the optical, it is in fact extremely difficult to determine the exact nature (sdB or sdO) of the hot subdwarf component. Based on the detection of the He~\textsc{ii} line at 4686 \AA, \citet{2012ASPC..452..233O} realized that the subdwarf component of PB 8783 must be hotter than previously believed. However, \citet{2012ASPC..452..233O} could not provide accurate estimates of the star's atmospheric parameters (he proposed $T_{\rm eff} \gtrsim$ 50 000 K and log $g \sim$ 6.05, without quoting uncertainties). The hypothesis of a sdO star was tested in preliminary asteroseismic analyses by \citet{2014ASPC..481..115V}, who concluded that although an sdB nature cannot be excluded, the pulsation modes observed in PB 8783 are indeed better recovered by an sdO model. PB 8783 is the brightest sdOV candidate and the one that has been most extensively observed in terms of light curves. If its atmospheric parameters can be constrained, and its sdO nature confirmed, PB 8783 will become a prime target for performing the first seismic analysis of an sdO star and thereby probing the properties of this evolved evolutionary stage.

In this proceedings devoted to PB 8783, we present in Sect. \ref{spectro} new high-resolution spectroscopic observations of the star that allow us to settle the question as to the nature of the hot subdwarf. Section \ref{sismo} presents the pulsation properties of PB 8783 and our improved asteroseismic modeling based on the updated atmospheric parameters. We stress that the spectroscopic and subsequent asteroseismic analyses presented here are work in progress. Section \ref{cc} presents our current conclusions and perspectives for coming work.

%\section{PB 8783: an old but still mysterious friend}

\section{PB 8783: sdB or sdO?}
\label{spectro}
\subsection{The context}
The main sequence F-type component of PB 8783 actually dominates any spectroscopic observations in optical, and a "de-pollution" procedure is needed to obtain the atmospheric parameters of the subdwarf. Various methods are available in spectroscopy, but none is easy to apply or fully satisfactory in the present case.

Our first attempts included subtraction of the composite spectra by template spectra of the F-type (F0 to F4), in order to obtain cleaned spectra of the hot subdwarf component (for details, see \citealt{2014ASPC..481..115V}). Analyzed in our usual way (grid of NLTE, H/He model atmospheres), these cleaned spectra gave effective temperature and surface gravity estimates typical for a subdwarf of the B type (sdB star). However, a small problem already attracted our attention, even in our low-resolution spectra: He I lines are almost absent (while predicted by the model), and strong He II lines (not predicted by the model) are present. This is more typical of a sdO star, i.e. a much higher effective temperature. %Furthermore, H10 to H16 lines are clearly visible in the spectrum.  \textbf{You would not see H10 to H16 in an sdO star... I don't know what you mean... if H10 to H16 are seen in the cleaned spectrum, I would say it is very likely from the companion, indicating that the clean spectrum is not so "clean". }

\subsection{This work: spectral analysis with XTgrid}
\begin{figure*}
	\centering
	\includegraphics[width=0.85\linewidth]{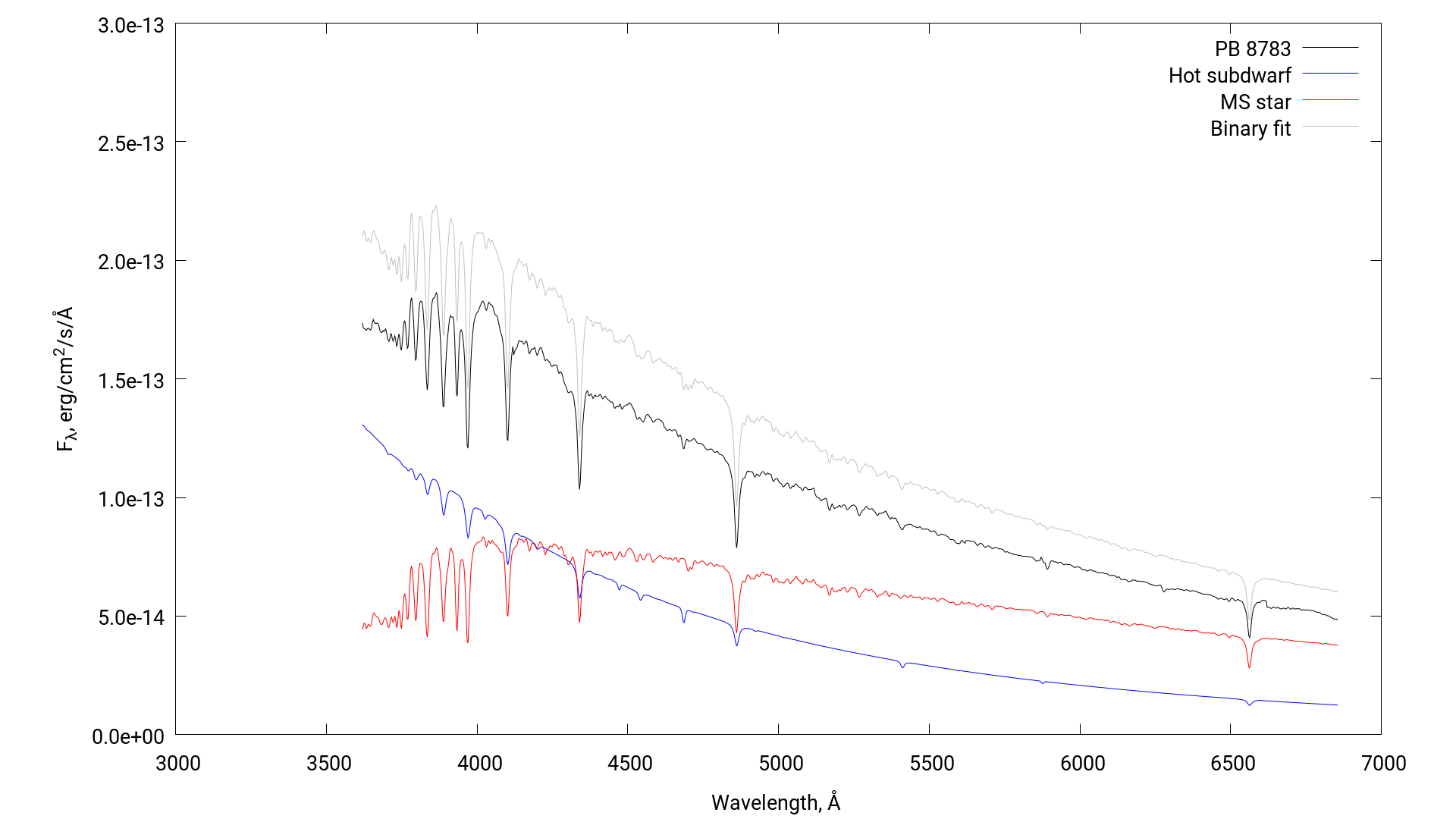}
	\caption{Low resolution PB 8783 spectrum (black) obtained with the Bok telescope. The synthetic composite model is the sum (grey) of the sdO type hot subdwarf model (blue) and the F0 type main sequence model (red). The composite model is shifted for clarity and fits the observation reasonably well.}
	\label{lowres}
\end{figure*}

\begin{figure*}
	\centering
	\includegraphics[width=0.85\linewidth]{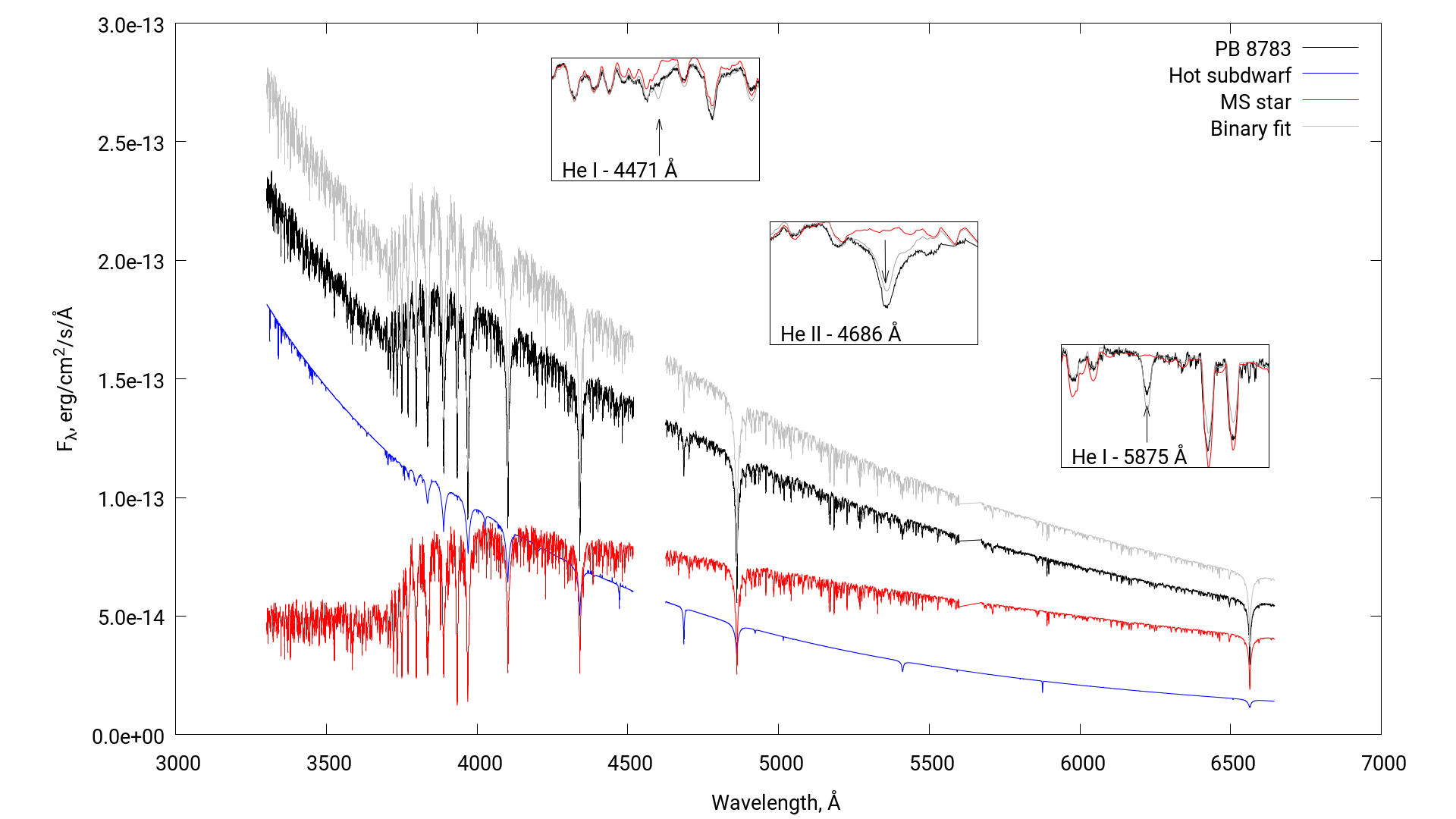}
	\caption{Same as Fig \ref{lowres}, but for the high resolution PB 8783 spectrum obtained with the VLT/UVES. The insets show the profiles of the He\,{\sc I} 4471 and 5875 \AA\ lines and the He\,{\sc II} 4686 \AA\ line.}
	\label{highres}
\end{figure*}

In this work, we analyze two very high-quality spectra: a low-resolution (9{\AA}), very high S/N ratio spectra obtained at the Bok telescope in Arizona (Fig. \ref{lowres}), and a high S/N, very high-resolution (about 0.1 {\AA}) spectra obtained in 2017 with UVES@VLT (Fig. \ref{highres}). The deconvolution procedure implemented in XTgrid \citep{2012MNRAS.427.2180N} uses a linear combination of synthetic sdO and F spectra and fits the observed composite spectrum with this combined spectrum. The contribution of each spectrum component ($T_{\rm eff}, \log g$, [Fe/H], $v \sin i$) is changed iteratively in a direction which improves the goodness of fit for as long as the $\chi^2$ decreases. The non-LTE sdO models are calculated with {\sc Tlusty/Synspec} \citep{2017arXiv170601859H} and the LTE models for the F-type companion are extracted from the BOSZ spectral library \citep{2017AJ....153..234B}.

The deconvolution of binaries consisting of very different types of components is usually an easy task. The distinct spectral features allow a precise separation of the stars even from a single observation. However, in the case of PB 8783 the F-type star dominates the flux above 4000 \AA\ and no clean lines of the sdO star are seen. The severe blending makes the Balmer series unreliable as an indicator of $T_{\rm eff}$/$\log{g}$. Metal absorption lines can help the spectral analysis and allow for a precise radial velocity and flux contribution measurement, but unfortunately none were found for the sdO star. Therefore we exploited the two sets of observations in tandem, using the low resolution Bok data to fit the slope of the composite continuum and the high resolution VLT spectrum to fit the profiles of the individual lines of both components. 

From this exercise we obtained $T_{\rm eff} \sim$ 52 000 K and log $g \sim$ 5.85 cm\,s$^{-2}$. The associated errors are difficult to estimate due to strong correlations. We tentatively estimate $\pm$ 3500 K and $\pm$ 0.15  cm\,s$^{-2}$. This clearly places the subdwarf component of PB 8783 in the sdO regime, as proposed by  \citet{2012ASPC..452..233O}.

Another intriguing result of the new analysis is the need for extra line broadening to consistently fit the helium lines of the sdO star. The broadening corresponds to a projected rotational velocity of about 50 km\,s${^{-1}}$ and it is likely due to pulsations (see Sect. \ref{sismo}). The broadening is large enough to smear weak metal lines into the continuum. 

\section{Pulsations in PB 8783}
\label{sismo}
\subsection{Photometric observations of PB 8783}
PB 8783 was the second subdwarf pulsator ever found \citep{1997MNRAS.285..645K}. Thanks to its brightness ($V=12.6$, one of the brightest subdwarf pulsators), it was frequently re-observed over the years to study its pulsations \citep[e.g.][]{1998MNRAS.296..296O,2000MNRAS.318..974J,2010Ap&SS.329..183V,2014ASPC..481...19F}. And in contrast to many other sdB pulsators, PB 8783 has exhibited a relatively stable pulsation spectrum over the years (Table \ref{supp17}). 

In this work, we used the frequencies extracted from our photometric campaign carried at Mt Bigelow (Arizona, USA) during 182.5h in 2007 \citep{2014ASPC..481...19F}. Details on the obtained pulsation spectrum can be found in \citet{2014ASPC..481..115V}, only the more important features are recalled here. We detected eight independent pulsation modes with amplitudes above 6$\sigma$ that were identified in almost all previous PB 8783 campaigns (Table \ref{supp17}). The Bigelow data also revealed 12 additional independent periods with amplitudes between 4.5 and 6$\sigma$ (we will not consider modes with lower amplitudes for the seismic modeling here). Finally, five of the eight basic pulsation modes exhibit some form of multiplet structure: 4 peaks around 116.43 s (i.e. an $l \geq$ 2 mode), a doublet at 122.68 s (at least $l = 1$), 4 peaks including the dominant mode at 123.63 s (therefore $l \geq 2$), 7 distinct peaks around 127.04 s (which favors $l = 4$), and a beautiful complete quintuplet at 134.17 s ($l \geq 2$, but most probably $l = 2$). These constraints on the mode identification will be used in the seismic analyses.
These multiplet structures are most likely caused by the rotation of the star in principle allow us to access the internal rotation of the subdwarf (see, for example, \citealt{2016ApJS..223...10G}). However, this has not yet been explored in detail. 

  \begin{table*}[!ht]
 \begin{center}
 {\small \begin{tabular}{c c c c c}\hline
{O'Donoghue (1998)}&{Jeffery \& Pollacco (2000)}&{Vuckovic et al. (2010)}&{Fontaine et al. (2014)}&{Rank}\\
 \hline 
 94.133&94.118&94.13&94.165&$f_9$\\
 94.454&...&...&94.452&${f_{10}}$\\
 116.418&116.809&116.417&<116.43>&${f_{11}}$\\
 <122.678>&122.835&122.692&<122.680>&${f_2}$\\
 123.578&...&123.692&<123.630>&${f_1}$,${f_5}$\\
 127.060&127.275&126.983&<127.044>&${f_4}$\\
 <134.165>&134.120&134.187&<134.169>&${f_3}$,${f_7}$\\
 136.269&136.258&...&136.273&${f_{12}}$\\
 \hline
 \end{tabular}}
 \end{center}
 \caption{Pulsation periods from the main photometric observations of PB 8783. Brackets <> indicates observed multiplets.}\label{supp17}
 \end{table*} 

\subsection{Models and Methods for sdB and sdO asteroseismic modeling}
\label{method}
Second-generation (2G) models are used to model PB 8783 in terms of an sdO, post-EHB pulsator. Our more recent 3G \citep{2013A&A...553A..97V} and 4G \citep{2018OAst...27...44V} models are complete structures only suitable for modelling core-He burning objects. In contrast, the 2G models are static envelope structures (the center is modeled as a hard ball), well suited to accurately compute p-mode pulsation periods \citep{2002ApJS..139..487C}. They incorporate the nonuniform iron abundances obtained by equilibrium between radiative levitation and gravitational settling, up to $T_{\rm eff} =$ 70 000 K. The input parameters of 2G models are the effective temperature $T_{\rm eff}$, the surface gravity log $g$, the total mass of the star $M_*$, and the mass contained in the H-rich envelope $\log (M_{\rm env}/M_*)\sim D(H)$.

The forward modeling approach developed to perform objective asteroseismic modeling of subdwarf pulsators has been described in detail in \citet{2008A&A...489..377C}. We fit directly and simultaneously the eight independent pulsation periods $P_{obs}$ of PB 8783 with the theoretical $P_{th}$ calculated from subdwarf models, in order to minimize the merit function defined by
\begin{equation}
S^2 (a_1,a_2,...,a_n) = \sum_{i=1}^{N_{obs}} \left( \frac{P_{obs,i} - P_{th,i}}{\sigma_i} \right)^2 \;\;\;.
\end{equation}
The $a_i$ are the parameters of the stellar models, $N_{\rm obs}$ is the number of observed independent periodicities, and $\sigma_i$ represents the weight of each pair \{$P_{obs}$,$P_{th}$\}. We chose $\sigma=1$ for each of them. The method performs a double-optimization procedure in order to find the minima of the merit function that constitute potential asteroseismic solutions, while adhering to external constraints from spectroscopy and from mode identification. We developed the efficient genetic algorithm LUCY that performs this optimization procedure by growing a random population of potential solutions from generation to generation \citep{2008A&A...489..377C}.

\subsection{Asteroseismic modeling of PB 8783}

The optimization procedure is launched in a vast parameter space where hot subdwarfs are found: $0.3\leq M_*/M_{\odot}\leq 0.7$, $-10.0\leq D(H) \leq -2.20$ \citep{2002MNRAS.336..449H,2003MNRAS.341..669H}. The surface gravity $\log{g}$ is searched between 5.7 and 6.1 (see Sect. \ref{spectro}). The effective temperature is fixed to its spectroscopic value, since H/He lines are much more sensitive to the effective temperature than the p-modes are \citep{2005A&A...437..575C}. The main difference to the preliminary seismic analysis presented in \citet{2014ASPC..481..115V} is that the parameter space is now much more exhaustively explored, with increased computational capacities. While we previously searched for solutions with a population of 100 individuals over 100 generations, we now typically grow 200 individuals over 500 generations. We also developed the computation of Probability Density Functions (PDFs) for the parameters to estimate realistic uncertainties \citep{2013A&A...553A..97V}, rather than focusing on the optimal solution and its surroundings (in the preliminary seismic analysis of \citealt{2014ASPC..481..115V}, we did not compute the uncertainties).

\begin{figure}
	\centering
	\includegraphics[width=1.0\linewidth]{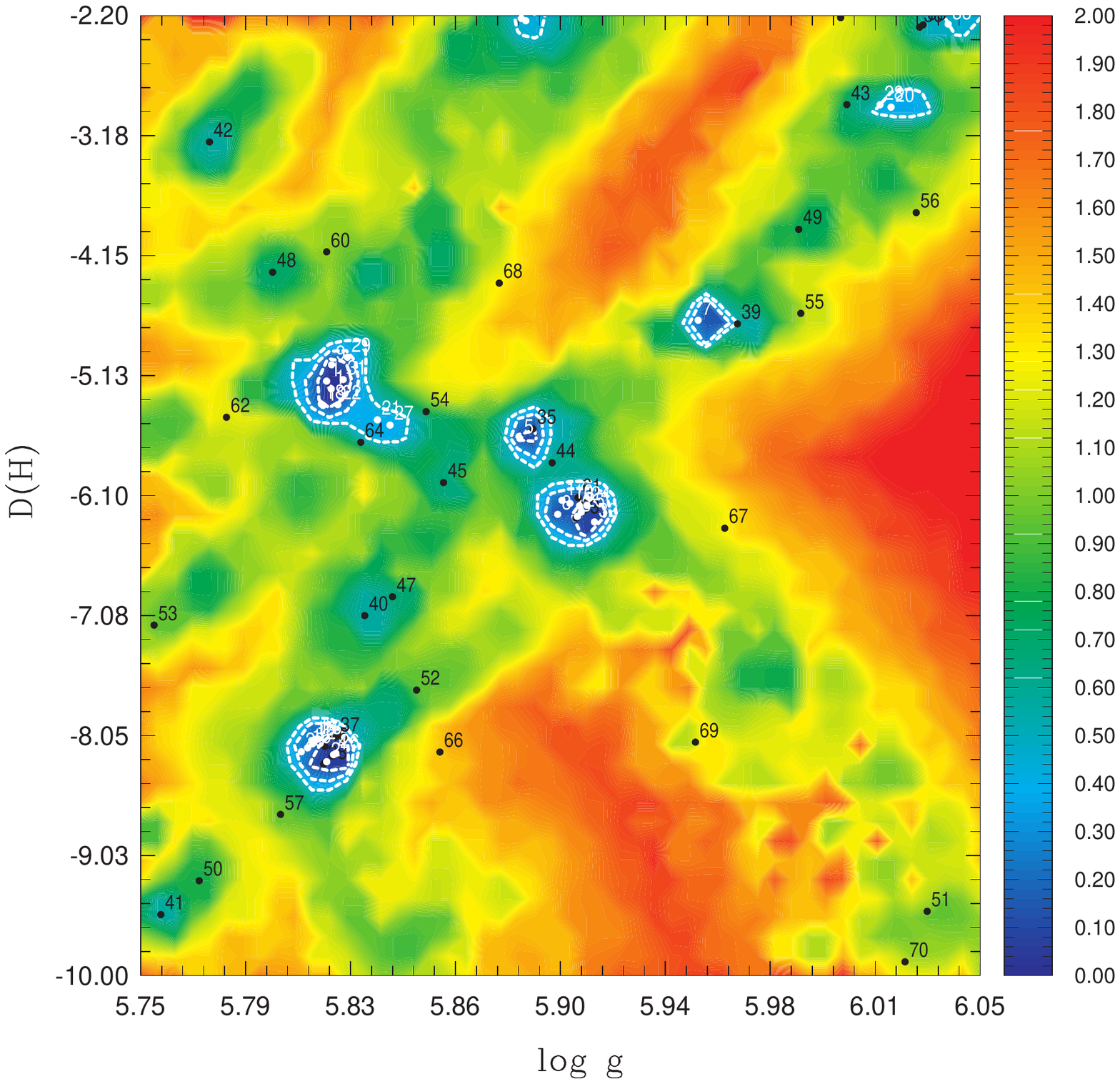}
	\caption{Map of the projected merit function $S^2$ (on a logarithmic scale) in the $\log{g} - D(H)$ plane. Best-fit models are identified by their rank (1 for the lowest merit function, etc). White contours show regions where the period fits have $S^2$ values within the 1$\sigma$, 2$\sigma$ and 3$\sigma$ confidence levels relative to the best-fit solution.}
	\label{map}
\end{figure}

Fig. \ref{histos} presents the PDFs obtained for the 3 parameters explored here: $M_*$, $\log{g}$, and $D(H)$. It is evident that we obtained several seismic solutions, with similar minima for their merit functions (this is particularly obvious on the map $\log{g} - D(H)$ on Fig. \ref{map}). These minima have period fits of $\overline{\Delta X/X} \sim 0.4\%$, which corresponds to $\Delta P \sim$ 0.4 s. Only the stellar mass seems relatively well constrained, with 0.42 $\pm$ 0.03 $M_{\odot}$ (Fig. \ref{histos}). Somewhat surprisingly, the mode identification revealed that while the shortest, low-degree ($l=0,1$) pulsations are associated with p-modes, the longest and higher degree ($l=2,4$) pulsations are associated with mixed modes. For example, the quintuplet at 134.17s has $(l,k)=(2,-8)$.

This multiplicity of asteroseismic solutions is partly explained by the relatively poor constraint on surface gravity $\log{g}$ from spectroscopy. We also think that the presence of mixed modes reveals the limits of using 2G envelope models (where the central parts are missing) to model sdO, post-EHB stars that are likely in a phase of He-shell burning. 

\begin{figure}
	\centering
	\includegraphics[width=0.85\linewidth]{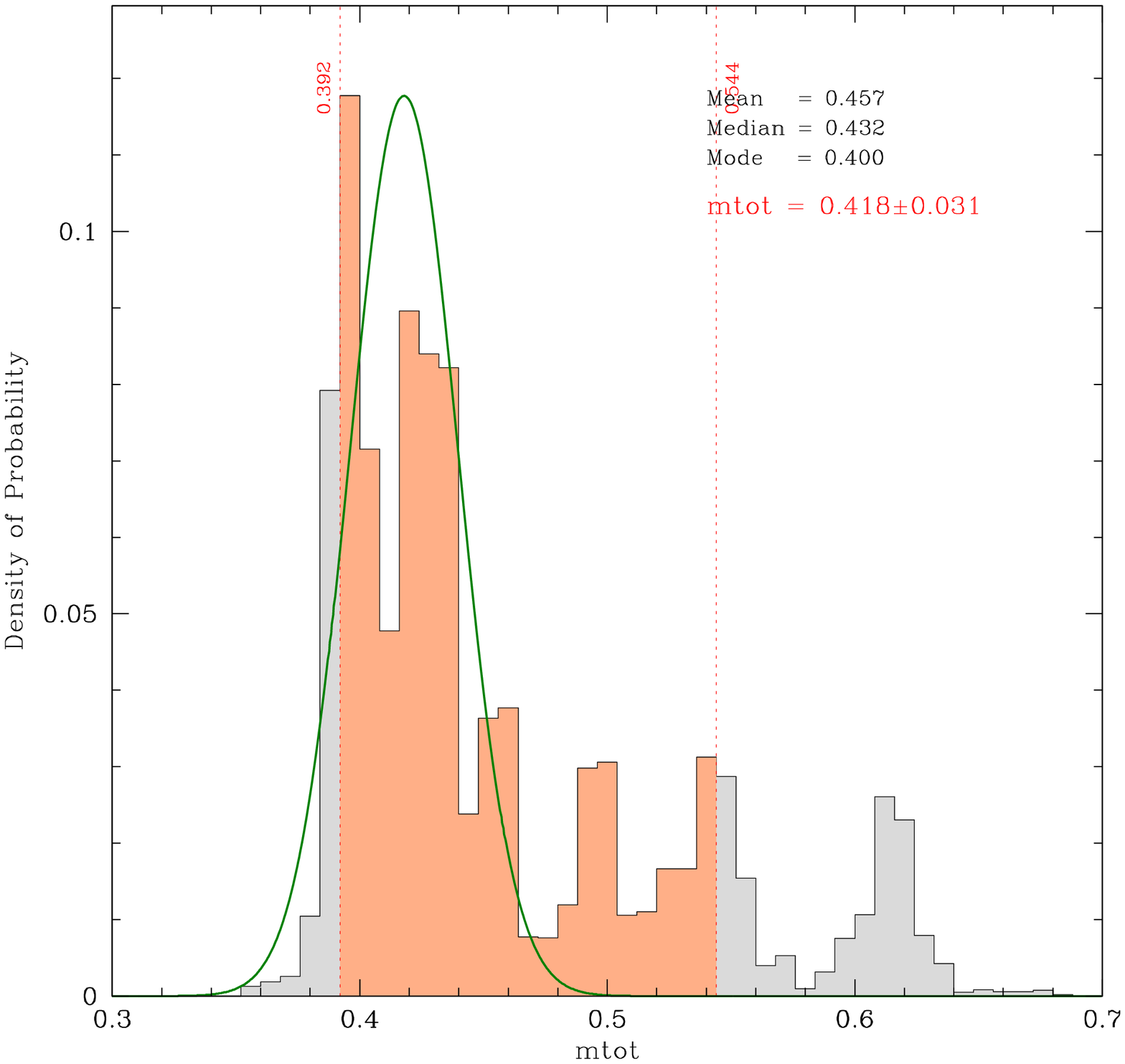}
	\includegraphics[width=0.85\linewidth]{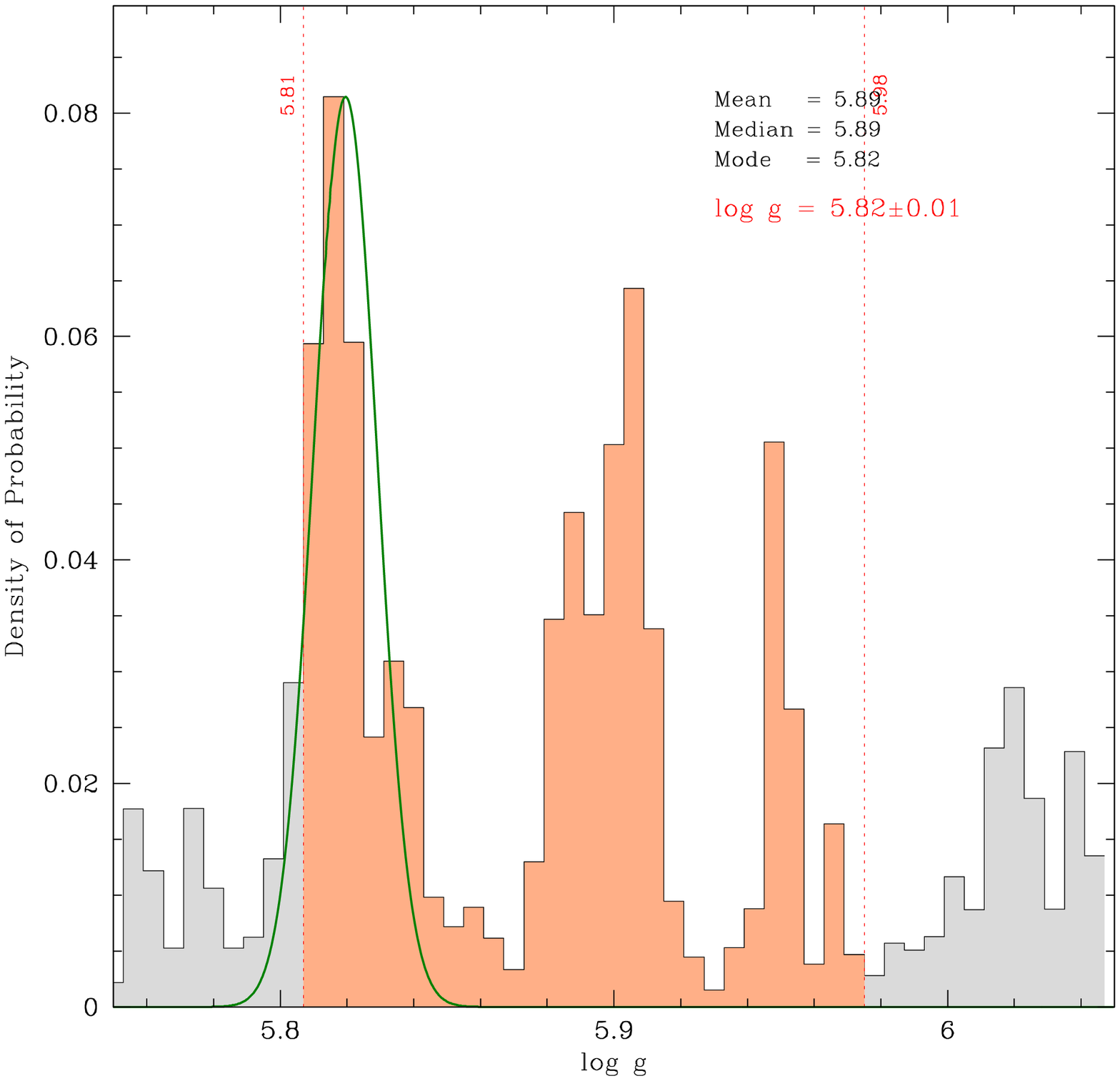}
	\includegraphics[width=0.85\linewidth]{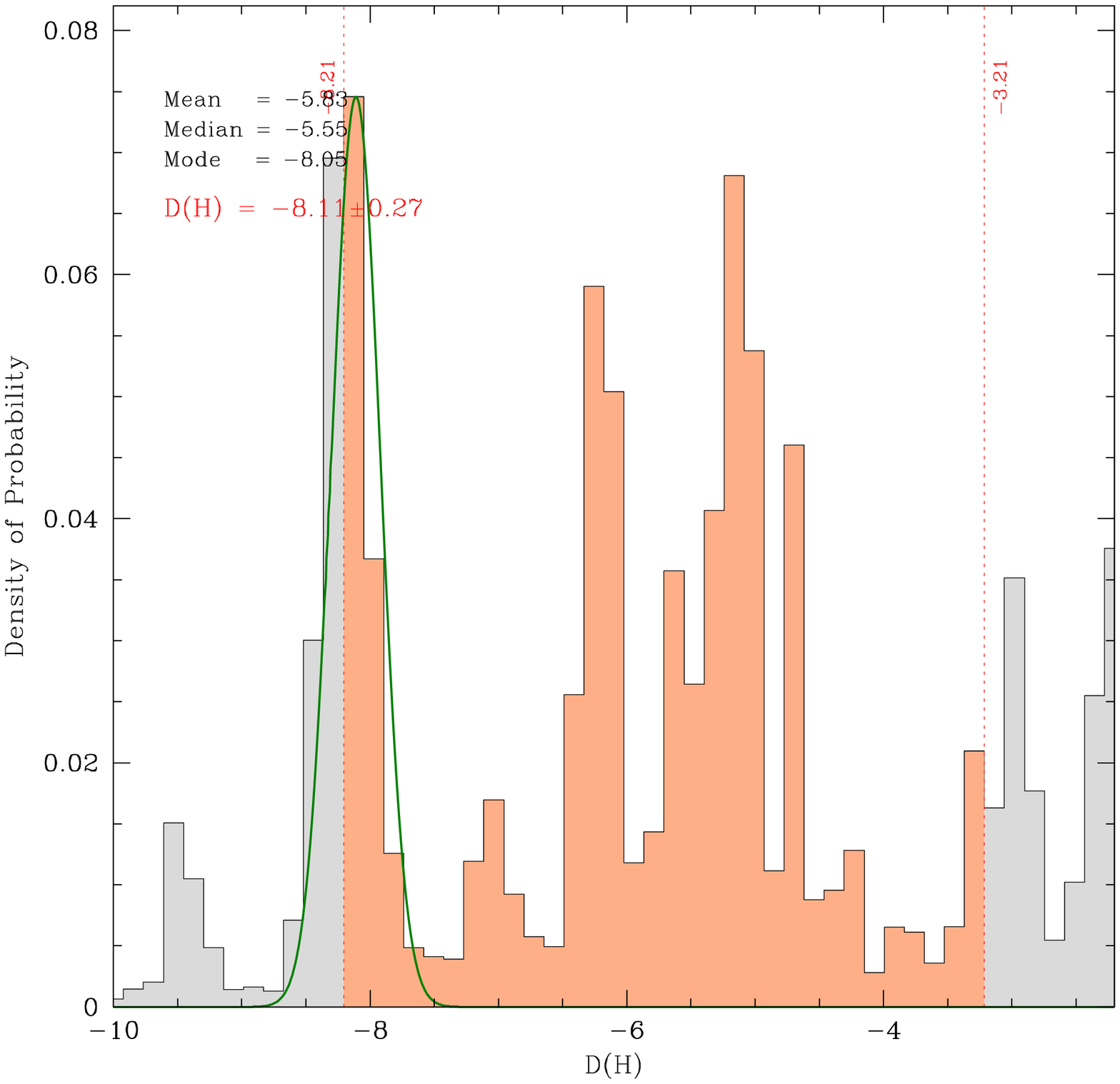}
	\caption{Probability density functions derived from asteroseismology for the 3 parameters of the 2G models: $M_*$, $\log{g}$, and $D(H) \equiv \log q(H)$. The red-hatched region between the two vertical solid red lines gives the $1\sigma$ range containing $68.3\%$ of the distribution. \label{histos}}
	\label{histos}
\end{figure}

\section{Conclusions and Prospects}
\label{cc}
PB 8783 is an old but mysterious friend, consisting of a pulsating subdwarf and a F-type main sequence companion. After many years (1997-2012) of being identified as an sdBV$_r$ pulsator, followed by years (2012-2018) of uncertainty about the exact nature of the subdwarf component (sdB or sdO), we finally obtained very high-quality spectroscopy to resolve the issue: the subdwarf pulsator in PB 8783 is definitely an sdO star,  with $T_{\rm eff} \sim$ 52 000 K and log $g \sim$ 5.85 cm\,s$^{-2}$. 

This is already an important conclusion, on the one hand because PB 8783 is one of the best observed subdwarf pulsators, and on the other hand because the sdO pulsators are rare and extremely difficult to study, both from a photometric and spectroscopic point of view, in the field as well as GCs. From this point of view, we can say that PB 8783 is the first sdO star identified as suitable for asteroseismic modeling.  

However, we still face two challenges. The first is that the deconvolution procedure to disentangle the contributions of the two components is particularly difficult, and strong correlations exist between spectroscopic parameters. Precise atmospheric parameters are therefore very hard to obtain. The second issue is that we have at our disposal only the 2G models to model sdO stars, which are post-EHB stars most likely in a He-shell burning phase (3G and 4G models are by construction limited to core-He burning objects). These 2G envelope models, where the central parts are missing, are probably not suitable for quantitatively modelling the pulsation periods of PB 8783, as evidenced by the fact that some of the observed periods are associated with mixed modes in our fits. 

To progress on the spectroscopic side, space-based UV spectroscopy would be helpful, however such observations are notoriously difficult to obtain. To progress on the modeling side, "special" complete models for post-EHB stars must be built, which is not a trivial task in our parameterized approach. The new GAIA DR3 distance and orbital motion resolution will definitely be of help for constraining our results. These are the avenues to progress on our understanding of PB 8783 in the coming years, and to make it the first sdO star modeled by asteroseismology, unveiling the properties of this advanced evolutionary stage. 

\section*{Acknowledgments}

Based on observations obtained at the Very Large Telescope (VLT) of the European Southern Observatory, Paranal, Chile (ESO Programme ID 099.D-0226(A)). 

\bibliographystyle{phostproc}
\bibliography{vangrootel.bib}

\end{document}